\documentclass[aps,prb,reprint,superscriptaddress,showpacs]{revtex4-1}

\usepackage{graphicx}
\usepackage{dcolumn}
\usepackage{bm}

\begin{document}


\title{Interlayer interaction in multilayer CoPt/Co structures}


\author{E. S. Demidov}
\affiliation{University of Nizhny Novgorod, 23 Prospekt Gagarina, 603950, Nizhny Novgorod, Russia}
\author{N. S. Gusev}
\affiliation{Institute for Physics of Microstructures RAS, GSP-105, 603950, Nizhny Novgorod, Russia}
\author{L.  I. Budarin}
\affiliation{University of Nizhny Novgorod, 23 Prospekt Gagarina, 603950, Nizhny Novgorod, Russia}
\author{E. A. Karashtin}
\author{V. L. Mironov}
\author{A. A. Fraerman}
\affiliation{University of Nizhny Novgorod, 23 Prospekt Gagarina, 603950, Nizhny Novgorod, Russia}
\affiliation{Institute for Physics of Microstructures RAS, GSP-105, 603950, Nizhny Novgorod, Russia}


\date{\today}

\begin{abstract}
We report a study of interlayer exchange interaction in multilayer CoPt/Co structures consisting of periodic CoPt multilayer film with an ``easy axis'' anisotropy and thick Co layer with an ``easy plane'' anisotropy separated by Pt spacer with variable thickness. The magnetooptical Kerr effect (MOKE) and ferromagnetic resonance (FMR) measurements show up the essentially non-collinear state of magnetic moments of the layers and strong exchange coupling between CoPt and Co subsystems. The estimation of effective anisotropy and exchange coupling in a simple model based on the Landau-Lifshitz-Gilbert equation describing magnetization dynamics was performed.
\end{abstract}

\pacs{75.30.Et, 75.50.Cc, 75.70.Cn}

\maketitle


\section{Introduction\label{Intro}}
Thin magnetic films and multilayer structures with ``easy axis'' anisotropy that have perpendicular magnetization are the subject of intensive research driven by promising applications in modern magnetic data storage systems. In recent years much attention has been paid to multilayer systems based on thin Co layers separated by noble metal spacers Me (such as Pt, Pd, Ni, Au) \cite{100Sbiaa,1Carcia,101Carcia,102Daalderop,103Broeder}. Perpendicular anisotropy in these systems is a characteristic property of the Co/Me interface \cite{1Carcia}. If Co layer thickness is less than a critical value ($\sim 1\,nm$) surface anisotropy exceeds shape anisotropy, and the multilayer CoMe system is perpendicularly magnetized. This property of CoMe multilayers considerably expands the opportunities in development of magneto-optical and magnetoresistive devices with lateral/vertical architecture \cite{2Stamps,3Zeper,4Kugler}.

Over the last few years certain attention has been focused on the properties of exchange coupled systems with distinct anisotropy directions \cite{20Tryputen,106Nguyen,105Tacchi,5Bollero,6Bollero,7Gong,8Nagaosa,9Sun,13Miao,10Sapozhnikov,11Sapozhnikov,12Fraerman}. In particular, CoPt/F structures that consist of CoPt multilayer film with ``easy axis'' anisotropy and ferromagnetic layer (F) with ``easy plane'' anisotropy have been proposed to control hysteresis loop shift (exchange bias) of magnetic system \cite{5Bollero,6Bollero,7Gong}. Besides, patterned CoPt/F structures were recently used to create artificial magnetic skyrmion states \cite{8Nagaosa,9Sun,13Miao,104Liu,10Sapozhnikov,11Sapozhnikov,12Fraerman}. Note that the effects of exchange bias and skyrmion nucleation are based on the existence of strong exchange interaction both between Co layers in CoPt film and between CoPt film and ferromagnetic layer F. However, as far as we know, there are no direct experimental studies of exchange coupling in such structures. The aim of this work is the study of interlayer exchange interaction in multilayer CoPt/Co structures by FMR method.

\section{Experimental Methods and Results\label{Experiment}}
\begin{figure}[b]
\includegraphics[width=2.2in, keepaspectratio=true]{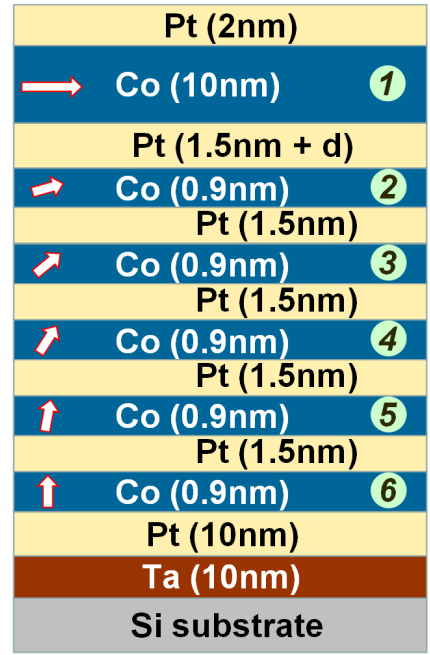}
\caption{\label{Fig_1} (Color online) A schematic drawing of CoPt/Pt(d)/Co multilayer structure. Numbers of layers are shown on the right side; arrows on the left side demonstrate magnetization distribution.
}
\end{figure}

Two [Co(0.9~nm)Pt(1.5~nm)]$_5$/Pt(d)/Co(10~nm) structures (further denoted as CoPt/Pt(d)/Co) with different spacer thickness $d = 0$ and $d = 1.5\,nm$ (see Fig.~\ref{Fig_1}) were grown by dc magnetron sputtering on a Si substrate with Ta(10~nm) and Pt(10~nm) underlayers \cite{12Fraerman}.
In addition, a multilayer [Co(0.9~nm)Pt(1.5~nm)]$_5$ structure and a Co(10~nm) film was grown separately on similar substrates. Thickness of the layers was determined with Bruker diffractometer (wavelength $\lambda = 0.154\,nm$) by the small-angle X-ray scattering method. Magnetization curves of the samples were measured by polar MOKE (He-Ne laser at $628\,nm$ wavelength). The residual sample domain structure was studied using vacuum magnetic force microscope (MFM) ``Solver-HV'' (NT-MDT Company). Standard NSG-11 cantilever with Co coating was used as an MFM probe. The measurements were performed in the constant-height mode with MFM contrast proportional to the phase shift of cantilever oscillations under the gradient of magnetic force \cite{22Thiaville,23Mironov}.

The FMR measurements were performed with Bruker EMX Plus-10/12 spectrometer equipped by a dc magnet with field $H$ up to $15\,kOe$. Polarized microwave magnetic field $\mathbf{h}$ at $9.8\,GHz$ frequency ($TE_{011}$ mode of resonant cavity) was perpendicular to the zero-frequency magnetic field $\mathbf{H}$. The samples were driven through resonance by sweeping the magnitude of magnetic field $H$. Angular dependences of resonant field position $H_r$ were investigated by rotation of the sample aroung the axis parallel to the direction of magnetic component of microwave field $\mathbf{h}$.

\begin{figure}[t]
\includegraphics[width=2.52in, keepaspectratio=true]{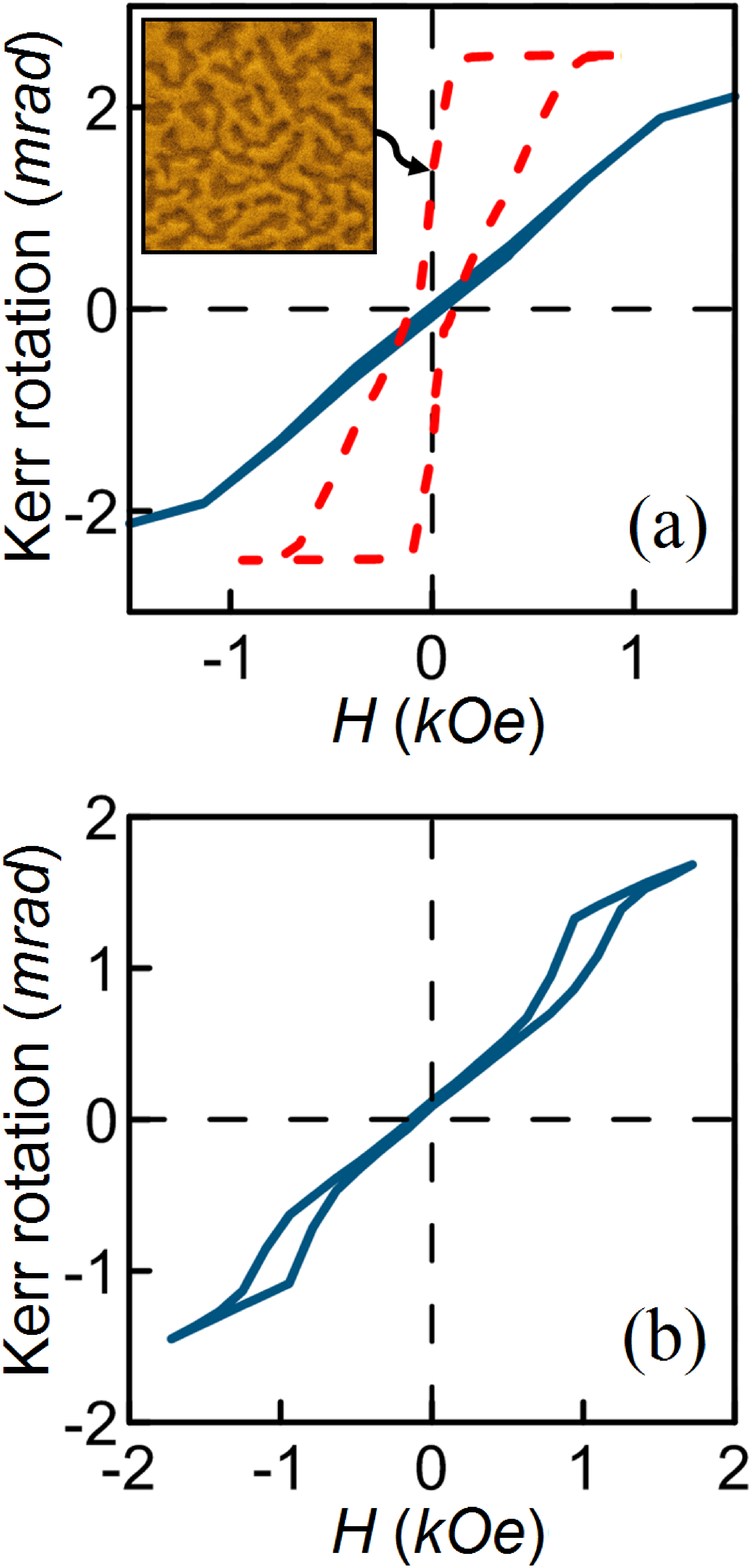}
\caption{\label{Fig_2} (Color online) MOKE magnetization curves of the samples. (a) The hysteresis loop of CoPt/Pt(0)/Co (solid line; blue online) and [Co(0.9~nm)Pt(1.5~nm)]$_5$ (dashed line; red online). The inset represents MFM image of the residual state of [Co(0.9~nm)Pt(1.5~nm)]$_5$ sample, frame size is $3\,\mu m$. (b) The hysteresis loop of CoPt/Pt(1.5~nm)/Co.}
\end{figure}

MOKE curves for CoPt/Pt(0)/Co and CoPt/Pt(1.5~nm)/Co are shown in Fig.~\ref{Fig_2}(a) and Fig.~\ref{Fig_2}(b) respectively.
The magnetization curve of [Co(0.9~nm)Pt(1.5~nm)]$_5$ sample is shown by dashed line in Fig.~\ref{Fig_2}(a). Note that remaining magnetization is less than magnetization in saturation due to splitting into domains which is confirmed by MFM measurements (inset in Fig.~\ref{Fig_2}(a)). The magnetization curve of CoPt/Pt(0)/Co sample is hysteresisless and is essentially different compared to the curve of [Co(0.9~nm)Pt(1.5~nm)]$_5$ sample. In contrast, the magnetization curve of CoPt/Pt(1.5~nm)/Co sample, although having distinctions from that of CoPt without a thick Co capper, keeps hysteresis, which is attributed to small influence of thick Co layer on CoPt subsystem. Significant change of magnetic properties of the multilayer CoPt structure caused by the thick cobalt film in CoPt/Pt(0)/Co sample is supposed to appear due to interaction of these subsystems through a platinum spacer. In order to study this effect in detail we used the FMR method.

\begin{figure}[t]
\includegraphics[width=2.58in, keepaspectratio=true]{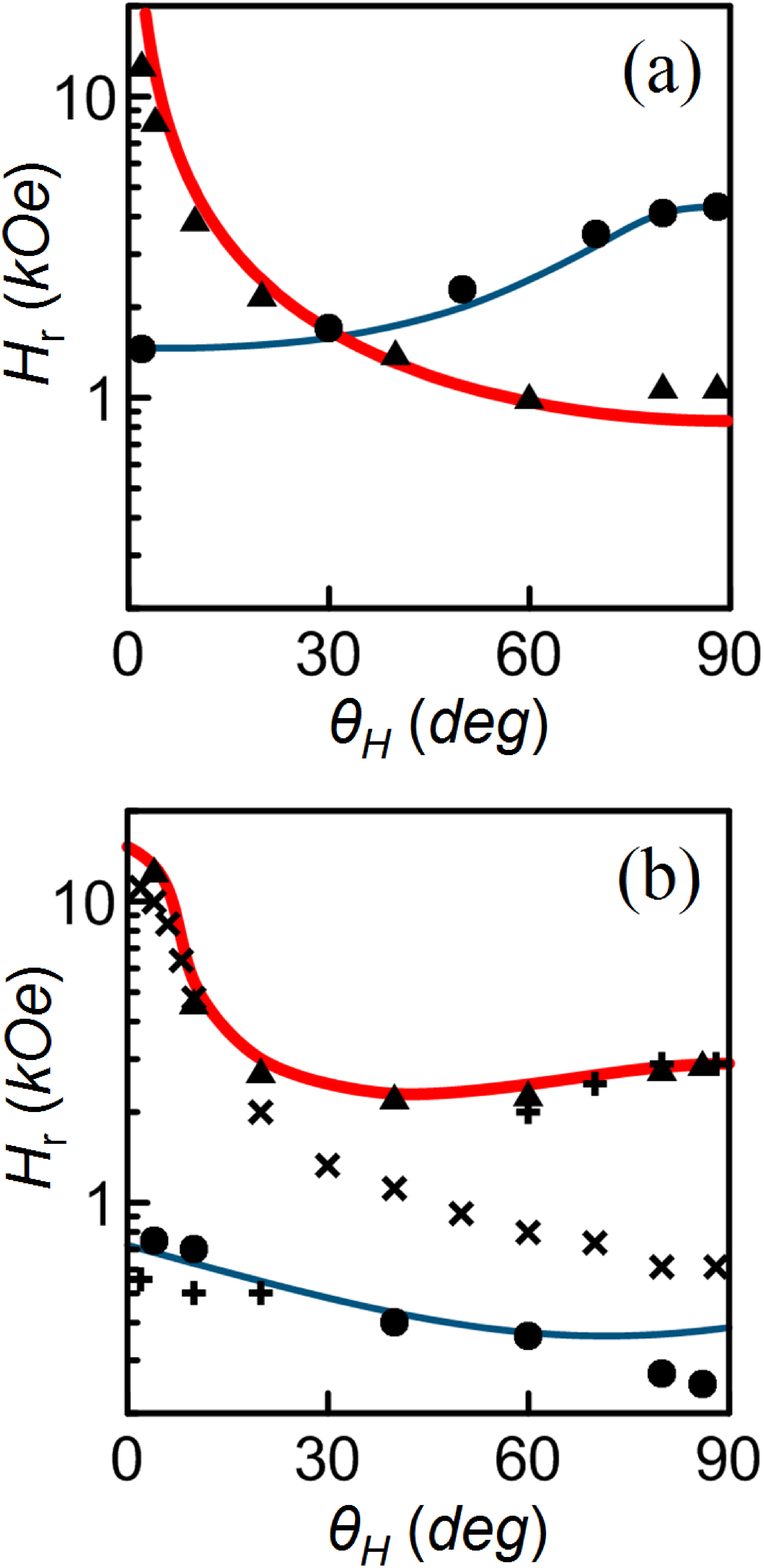}
\caption{\label{Fig_3} (Color online) Angular dependences of resonant field position $H_r (\theta_H)$ (a) for Co layer (experimental black triangles and calculated thick line (red online)) and for [Co(0.9~nm)Pt(1.5~nm)]$_5$ structure (experimental black circles and calculated thin line (blue online)); (b) for CoPt/Pt(0)/Co sample, acoustic mode (experimental black triangles and calculated thick line (red online)) and optical mode (experimental black circles and calculated thin line (blue online)). Crosses and pluses in (b) are experimental data for two modes in CoPt/Pt(1.5~nm)/Co~ sample.}
\end{figure}

Dependence of resonant magnetic field $H_r$ on the angle $\theta_H$ between $\mathbf{H}$ and the normal to layers plane for a Co(10~nm) film and a multilayer [Co(0.9~nm)Pt(1.5~nm)]$_5$ structure is shown in Fig.~\ref{Fig_3}(a).
Note that all investigated structures show angular resonant field dependences symmetric with respect to the values of $\theta_H$ equal to multiples of $90^\circ$ and $180^\circ$ and are periodic with the $180^\circ$ period. This proves there is no observable lateral magnetic anisotropy. Therefore all experimental and calculated data on resonant fields are hereafter given for $\theta_H$ lying in the interval between $0$ and $90^\circ$ and the in-plane anisotropy is neglected in calculations.
Comparing resonant fields for separate Co(10~nm) film and [Co(0.9~nm)Pt(1.5~nm)]$_5$ structure (see Fig.~\ref{Fig_3}(a)) for the external field parallel and perpendicular to layers plane ($\theta_H = 90^\circ$ and $0$ correspondingly) we see that thick cobalt film has the ``easy plane'' anisotropy while CoPt multilayer structure is a magnetic system with the ``easy axis'' anisotropy (the axis is perpendicular to layers plane). This agrees with the magnetization curve and MFM measurements due to splitting into domains.

Angular dependences of resonant field for the CoPt/Pt(0)/Co sample with thin spacer ($1.5\,nm$ total) between CoPt and Co are shown in Fig.~\ref{Fig_3}(b). We observed large splitting for two branches of resonant oscillations, which is a direct evidence of strong interaction between subsystems. Different dependences are observed for a sample with thick Pt spacer ($3\,nm$ total, $d = 1.5\,nm$) between Co and CoPt (crosses and pluses in Fig.~\ref{Fig_3}(b)). In this case there are two almost independent subsystems. Simple estimations show that such strong dependence of $H_r (\theta_H)$ splitting on thickness of Pt spacer excludes the magnetostatic interaction mechanism \cite{14Schrag} and therefore we suggest the exchange interaction between Co and CoPt subsystems.

\section{Discussion\label{Disc}}
In order to estimate the value of exchange interaction in the CoPt/Pt(0)/Co sample we consider the following simple model. We suppose that magnetic moment $\mathbf{M}_i$ of i-th Co layer (see Fig.~\ref{Fig_1})  is constant throughout the layer; $\mathbf{M}_i$ lies in a plane formed by the vectors of normal and external magnetic field $\mathbf{H}$ and its orientation is determined by the angle $\theta_i$ with respect to the normal to layers. Saturation magnetization $M$ is the same for all layers and is supposed to be equal to the volume value \cite{15Babichev} $M \approx 1420\,G$. Effective anisotropy of each cobalt layer is determined by interplay of the uniaxial anisotropy and the demagnetizing factor. Assuming that value of uniaxial anisotropy depends only on the layer thickness we take equal anisotropy constant $K$ for layers 2 to 6 and a different constant $K_c$ for thick Co layer 1. We also use equal constants of exchange interaction between all neighbour cobalt layers taking into account that platinum interlayers between them have the same thickness in the considered structure. Then the expression for surface energy density can be written in the following form:
\begin{eqnarray}
E &=& \left(2 \pi M^2 - K_c\right) l_1 \cos^2 \theta_1 + \nonumber \\
 &+& \left(2 \pi M^2 - K\right) \sum_{i=2}^{6}{l_i \cos^2 \theta_i} - \label{Eq_1} \\
&-& \sum_{i=1}^{6}{l_i \mathbf{H} \cdot \mathbf{M}_i} - \frac{J}{M^2} \sum_{i=1}^{5}{\mathbf{M}_i \cdot \mathbf{M}_{i+1}}, \nonumber
\end{eqnarray}
where $l_i$ is the i-th layer thickness, $J$ is interlayer exchange interaction constant (per unit area).

Dynamics of magnetic moments is desctibed by a system of Landau-Lifshitz-Gilbert equations:
\begin{equation} \label{Eq_2}
\frac{\partial \mathbf{M}_i}{\partial t} = - \gamma_i \left[\mathbf{M}_i \times \mathbf{H}_i\right] + \frac{\alpha}{M} \left[\mathbf{M}_i \times \left[ \mathbf{M}_i \times \mathbf{H}_i \right] \right],
\end{equation}
where $\mathbf{H}_i = - \frac{\partial E}{\partial \mathbf{M}_i}$ is the effective magnetic field acting on the i-th ferromagnetic layer (determined by (\ref{Eq_1})), $\gamma_i$ is the gyromagnetic ratio that is slightly different for CoPt and Co due to different g-factor. The parameter $\alpha$ in (\ref{Eq_2}) is the damping constant, which generally determines the width of FMR lines. The dependence of resonant fields on system parameters and particularly on the value of exchange interaction constant $J$ are determined by the system (\ref{Eq_2}) linearized in the vicinity of equilibrium state ($\left[\mathbf{H}_i \times \mathbf{M}_i\right] = 0$). We acquire the best fit of experimental data with following parameters: $K = 1.36 \cdot 10^7\,erg/cm^3$; $K_c = 4 \cdot 10^6\,erg/cm^3$; $J = 2\,erg/cm^2; g_1 = 2; g_{2-6} = 2.07$ ($g_i$ is a g-factor that determines $\gamma_i$). Calculated dependences of resonant fields on the angle $\theta_H$ are shown with solid lines in Fig.~\ref{Fig_3}. The effective anisotropy of cobalt layers in multilayer CoPt structure ($i = 2..6$) $\left(2 \pi M^2 - K\right) \approx -1.3 \cdot 10^6\,erg/cm^3$ is negative (``easy axis'' type), the effective anisotropy of thick Co film $\left(2 \pi M^2 - K_c\right) \approx 8.7 \cdot 10^6\,erg/cm^3$ is positive (``easy plane'' type). Exchange interaction between subsystems with different anisotropy types leads to collective magnetization oscillations. The calculations show that the high-field resonant branch stands for cophased oscillations of magnetic moments of thick Co film and layers of the multilayer CoPt structure while the low-field branch is characterized by antiphased oscillations. This corresponds to ferromagnetic type of exchange interaction between the subsystems ($J > 0$).

\begin{figure}[t]
\includegraphics[width=2.6in, keepaspectratio=true]{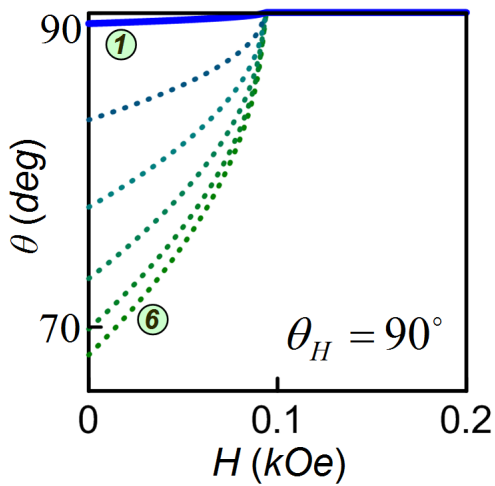}
\caption{\label{Fig_4} (Color online) Dependences of equilibrium angles of magnetic moments $\theta_i$ of layers on external magnetic field applied along the sample plane. Thick curve corresponds to the Co(10~nm) layer, thin dotted curves correspond to CoPt layers in order of increasing distance from thick Co layer.}
\end{figure}

The exchange interaction leads to non-collinear magnetization distribution characterized by magnetic moments lying neither in layers plane nor perpendicular to them (schematically shown in Fig.~\ref{Fig_1}). Results of calculations of equilibrium magnetic states under applied in-plane magnetic field are shown in Fig.~\ref{Fig_4}.
We see that the angle between the magnetic moment and the normal to sample increases with the number of the layer. This distribution is very similar to that demonstrated by thin films with different types of surface and volume anisotropy \cite{16Mills,17Allenspach,18Bochi,19Hu}.

\section{Conclusion\label{Sum}}
We performed the investigation of exchange interaction in a multilayer CoPt/Co structure that consists of a thick Co layer with ``easy plane'' anisotropy and a periodic CoPt film with ``easy axis'' anisotropy by ferromagnetic resonance method. It was demonstrated that in case of thin platinum spacer ($1.5\,nm$ total) strong exchange interaction between Co and CoPt subsystems exists, while for a thick spacer ($3\,nm$ total) the interaction is practically absent. The exchange interaction between subsystems gives an opportunity to create artificial structures with non-collinear magnetization distribution, such as a Neel-type magnetic structure in which magnetic moment has a component along the direction of its change. Particularly, these structures are of certain interest for the experimental observation of the flexo-magnetoelectric effect that was predicted in non-collinear systems \cite{21Baryakhtar}. To sum up, our investigations shed light on non-uniform multilayer magnetic systems with large value of magnetization gradient. Variation of interlayer spacing enables an effective control of non-collinear states and fine tuning of exchange interaction in such systems that is very important from the standpoint of practical applications.

\begin{acknowledgments}
The authors are thankful to O.L.~Ermolaeva for valuable discussions. 

Sample fabrication and MOKE investigations were supported by Russian Science Foundation (Grant No. 16-12-10340). MFM and FMR measurements were supported by Russian Science Foundation (Grant No. 16-12-10254). Theoretical calculations were supported by Russian Foundation for Basic Research. E.S.D. thanks the program ``Development of the Scientific Potential of Higher Education'', project no. 02V.49.21.0003 for Nizhny Novgorod State University.
\end{acknowledgments}

\bibliography{fmr}

\end{document}